\documentclass[12pt]{article}

\usepackage[pdftex]{graphicx}

\usepackage{epsfig,cite}

\usepackage {amsmath}

\usepackage{color}
\usepackage{rotating}
\usepackage{amssymb}

\usepackage{slashed}

\include{epsf}

\newcount\timecount

\newcount\hours \newcount\minutes  \newcount\temp \newcount\pmhours

\hours = \time

\divide\hours by 60

\temp = \hours

\multiply\temp by 60

\minutes = \time

\advance\minutes by -\temp

\def\hour{\the\hours}

\def\minute{\ifnum\minutes<10 0\the\minutes

            \else\the\minutes\fi}

\def\clock{

\ifnum\hours=0 12:\minute\ AM

\else\ifnum\hours<12 \hour:\minute\ AM

      \else\ifnum\hours=12 12:\minute\ PM

            \else\ifnum\hours>12

                 \pmhours=\hours

                 \advance\pmhours by -12

                 \the\pmhours:\minute\ PM

                 \fi

            \fi

      \fi

\fi

}

\def\monthname{\relax\ifcase\month 0/\or January\or February\or

   March\or April\or May\or June\or July\or August\or September\or

   October\or November\or December\else\number\month/\fi}

\def\bold#1{\setbox0=\hbox{$#1$}%

     \kern-.025em\copy0\kern-\wd0

     \kern.05em\copy0\kern-\wd0

     \kern-.025em\raise.0433em\box0 }

\def\beq{\begin{equation}}

\def\eeq{\end{equation}}

\def\ga{\mathrel{\raise.3ex\hbox{$>$\kern-.75em\lower1ex\hbox{$\sim$}}}}

\def\la{\mathrel{\raise.3ex\hbox{$<$\kern-.75em\lower1ex\hbox{$\sim$}}}}

\def\gev{{\rm \, Ge\kern-0.125em V}}

\def\tev{{\rm \, Te\kern-0.125em V}}

\def\gyr{{\rm \, G\kern-0.125em yr}}

\def\gappeq{\mathrel{\rlap {\raise.5ex\hbox{$>$}}

{\lower.5ex\hbox{$\sim$}}}}

\def\lappeq{\mathrel{\rlap{\raise.5ex\hbox{$<$}}

{\lower.5ex\hbox{$\sim$}}}}

\def\Toprel#1\over#2{\mathrel{\mathop{#2}\limits^{#1}}}

\def\m12{m_{1\!/2}}

\def\bea{\begin{eqnarray}}

\def\eea{\end{eqnarray}}

\def\beqar{\begin{eqnarray}}

\def\eeqar{\end{eqnarray}}

\def\m{{\cal m}}

\begin{document}

\begin{titlepage}

\pagestyle{empty}

\baselineskip=21pt


\rightline{KCL-PH-TH/2013-11, LCTS/2013-06, CERN-PH-TH/2013-057}

\vskip 0.7in

\begin{center}

{\large {\bf Wess-Zumino Inflation in Light of Planck}}

\end{center}

\begin{center}

\vskip 0.4in

{\bf Djuna~Croon}$^{1}$,
{\bf John~Ellis}$^{1,2}$
and {\bf Nick~E.~Mavromatos}$^{1,2}$

\vskip 0.2in

{\small {\it

$^1${Theoretical Particle Physics and Cosmology Group, Physics Department, \\
King's College London, London WC2R 2LS, UK}\\

$^2${TH Division, Physics Department, CERN, CH-1211 Geneva 23, Switzerland}\\

}}

\vskip 0.4in

{\bf Abstract}

\end{center}

\baselineskip=18pt \noindent


{
We discuss cosmological inflation in the minimal Wess-Zumino model with a single
massive chiral supermultiplet. With suitable parameters and assuming a plausible initial condition at the start of the
inflationary epoch, the model can yield scalar perturbations in the Cosmic Microwave Background (CMB)
of the correct strength with a spectral index
$n_s \sim 0.96$ and a tensor-to-scalar perturbation ratio $r < 0.1$, consistent with the Planck CMB data.
We also discuss the possibility of topological inflation within the Wess-Zumino model, and the possibility
of combining it with a seesaw model for neutrino masses. This would violate $R$-parity, but at such a
low rate that the lightest supersymmetric particle would have a lifetime long enough to constitute the
astrophysical cold dark matter.
}


\vfill

\leftline{
March 2013}

\end{titlepage}

\baselineskip=18pt


\section{Introduction and Summary}

There have been many discussions of single-field models of chaotic inflation based on
renormalizable polynomial potentials~\cite{encyclopedia}, i.e., combinations of $\phi^n : n \le 4$.
Prior to the Planck data on the Cosmic Microwave Background
(CMB)~\cite{Planck}, upper limits on the ratio $r$ of tensor and scalar
density perturbations and measurements of the scalar index $n_s$ from WMAP~\cite{wmap} and other CMB experiments already
disfavoured $\phi^4$ models quite strongly, and $\phi^2$ models were
marginal. This disfavouring of $\phi^n$ models with $n\ge 2$ has been reinforced by the
Planck data, which provide the strengthened upper limit $r < 0.11$ and constrain $n_s = 0.9603 \pm 0.0073$~\cite{Planck}.
Models with potentials of the form $\alpha \phi^2 + \beta \phi^4$ with
positive coefficients interpolate between pure $\phi^2$ and $\phi^4$
models and are therefore also disfavoured~\footnote{There has been interest in 
models with a linear potential $\propto \phi$~\cite{MSW,Hiranya}, and even in models with a fractional power
of $\phi$~\cite{SW}, though these can only be considered as effective models~\cite{encyclopedia}.}. For these and many other
reasons, attention has generally diffused to models with non-renormalizable
potentials and/or multiple fields, many of which are also excluded or disfavoured by the Planck CMB data~\cite{Planck}.

However, before abandoning renormalizable single-field models entirely,
we would like to advocate a particular example with attractive properties,
namely
\begin{equation}
V \; = \; A \phi^2 (v - \phi)^2 \, ,
\label{model}
\end{equation}
which has several interesting
aspects. For example, with reference to the title of this paper, it appears naturally
as the restriction of the minimal single-superfield Wess-Zumino model~\cite{WZ} 
characterized by the superpotential
\begin{equation}
W \; = \; \frac{\mu}{2} \Phi^2 - \frac{\lambda}{3} \Phi^3
\label{WZW}
\end{equation} 
to the real scalar component $\phi$ of the superfield $\Phi$~\footnote{Neither of the models
(\ref{model}, \ref{WZW}) seems to be considered in the recent review~\cite{encyclopedia}.}.
Another interesting feature of the model (\ref{model}) is that,
thanks to the two minima at $\phi = 0, v$ and the local maximum at $\phi = v/2$,
it leads to topological domain-wall inflation if $v \ga M_{Pl}$,
where $M_{Pl} \simeq 1.2 \times 10^{19}$~GeV is the Planck mass.
A third interesting feature of the model (\ref{model}) is that might be a viable extension of
the minimal supersymmetric seesaw model of neutrino masses with $\mu \ne 0$ and $\lambda = 0$, 
if one interprets $\Phi$ as a right-handed singlet neutrino superfield.
In this case one could envisage a scenario of chaotic sneutrino inflation
followed by leptogenesis during the subsequent reheating~\cite{ERY}. As we show below,
the simple model (\ref{model}) and its Wess-Zumino extension (\ref{WZW})
may overcome the disfavouring by the WMAP~\cite{wmap} and Planck~\cite{Planck} CMB data
of  chaotic inflationary models with monomial $\phi^n: n \ge 2$ potentials.

In this paper we first consider the minimal single-field model (\ref{model})
and discuss the conditions under which it can lead to acceptable chaotic
inflation in the slow-roll approximation. We show that the model yields enough e-folds of inflation if
the value of $v$ is large enough, typically $\gg M_{Pl}$, and that the
tensor-to-scalar ratio $r$ can be arbitrarily small in the limit where the 
initial value of the inflaton field $\phi_0 \to 1/2^-$. Thus this simple
single-field model is very consistent with the Planck CMB data~\cite{Planck}. We also note that the large value of
$v$ lies well within the range $\ga M_{Pl}$ where domain-wall inflation
is possible. In the case of the Wess-Zumino extension (\ref{WZW}) of
the minimal model, one may parametrize the complex scalar component
of $\Phi$ as $\phi e^{i \theta}$ and recover the simplified model (\ref{model})
in the limit $\theta \to 0$, identifying $A = \lambda^2$ and $v = \mu/\lambda$.
In this case, a secondary minimum at $\phi \ne 0$ appears only for
$\cos \theta > \sqrt{8/9}$, and is energetically disfavoured for $\cos \theta < 1$.
This suggests that the region of the minimum with $\phi = v$ would
generically be less populated than the region of the $\phi = 0$ minimum,
though this depends on aspects of the pre-inflationary dynamics that we
do not consider here. We conclude with some remarks about the possible
compatibility of the Wess-Zumino model (\ref{WZW}) with a supersymmetric seesaw model
of neutrino masses, pointing out that this would violate $R$-parity, though
not jeopardizing the possibility that the lightest supersymmetric particle
might provide the astrophysical cold dark matter.

\section{Basic Formulae}

For convenience in the following, we parameterize $\phi = x v$, and write
the effective potential obtained from (\ref{WZW}) in the form
\begin{equation}
V \; = \; \left| \frac{\partial W}{\partial \phi} \right|^2 \; = \; A v^4 ( x^4 - 2 \cos \theta x^3 + x^2) \, ,
\label{xparam}
\end{equation}
where, as already stated, we identify $A = \lambda^2$ and $v = \mu/\lambda$.
We recall that the measured magnitude of the primordial density perturbations
requires in the slow-roll approximation~\cite{encyclopedia}
\begin{equation}
\label{WMAPcon} 
\left(\frac{V}{\epsilon}\right)^{\frac{1}{4}} =  0.0275 \times M_{Pl} \, ,
\end{equation}
where the slow-roll parameter $\epsilon$ is given by~\cite{encyclopedia}
\begin{eqnarray} 
\epsilon \; = \; \frac{1}{2} M_{Pl}^2 \left( \frac{V'}{V} \right)^2 =  2 \frac{M_{Pl}^2}{v^2} \frac{1}{x^2}\, \Big[ 1 + \frac{x\, (x - {\rm cos}\theta )}{x^2 - 2 {\rm cos}\theta \, x + 1} \Big]^2 
\label{epsilon}
\end{eqnarray}
which in the limit ${\rm cos}\,\theta \rightarrow 1$, relevant for the single-field model (\ref{model}) becomes: 
\begin{equation}
\epsilon = 2 \frac{M_{Pl}^2}{v^2} \left[ \frac{(1 - 2 x)^2}{x^2 (1 - x)^2}\right] \, . 
\label{limittheta}
\end{equation}

The corresponding expressions for the other slow-roll parameters are~\cite{encyclopedia}
\begin{eqnarray} 
\eta \; = \; M_{Pl}^2 \left( \frac{V''}{V} \right)  =   2 \frac{M_{Pl}^2}{v^2} \left[ 1 + \frac{x\, ( 5 x - 4 {\rm cos}\, \theta )}{x^2 - 2 {\rm cos}\, \theta \, x + 1}\right] \, ,
\label{eta}
\end{eqnarray}
and
\begin{eqnarray} 
\xi \; = \; M_{Pl}^4 \left( \frac{V'V'''}{V^2} \right)  =  24\,  \frac{M_{Pl}^4}{v^4} \, \frac{(2x - {\rm cos}\, \theta) \, ( 2x^2 - 3\, {\rm cos}\, \theta \, x + 1)}{x^3\, (x^2 - 2\, {\rm cos}\theta \, x + 1)^2}~,
\label{xi}
\end{eqnarray}
which in the limit ${\rm cos}\, \theta \rightarrow 1 $ become:
\begin{eqnarray} 
\eta \; = \;  2 \frac{M_{Pl}^2}{v^2} \left[ \frac{(1 -6 x + 6 x^2)}{x^2 (1 - x)^2}\right] \, ,
\label{etalimit}
\end{eqnarray}
and 
\begin{eqnarray} 
\xi \; = \;   24\,  \frac{M_{Pl}^4}{v^4} \, \frac{(2x - 1 ) \, (2x^2 - 3x + 1)}{x^3\, (1 - x)^2}~.
\label{xilimit}
\end{eqnarray}
One can express the scalar spectral index in terms of the slow-roll parameters as~\cite{encyclopedia}
\begin{equation} 
n_s \; = \; 1 - 6 \epsilon + 2 \eta \, ,
\label{ns}
\end{equation}
and the tensor-to-scalar ratio as
\begin{equation} 
r \; = \; 16 \epsilon \, .
\end{equation}
Finally, the number of e-folds is given by~\cite{encyclopedia}
\begin{equation} 
\label{efoldscon} 
N \; = \; \frac{v^2}{M_{Pl}^2} \int^{x_e}_{x_i} \left( \frac{V}{V'} \right) dx \, ,
\end{equation}
where $x_{e,i}$ are the values of $x$ at the end and beginning of the
inflationary epoch. Assuming that
$x_e \ll 1$, we find that
\begin{eqnarray}
N = \frac{v^2}{16 M_{Pl}^2} \left[ - \ln (1 - 2 x_i) - 2 x_i + 2 x_i^2 \right]
\label{Nformula}
 \end{eqnarray}
in the limit ${\rm cos}\, \theta \rightarrow 1 $, and we expect that $40 \la N \la 70$.

For completeness, we also consider the 
running of the spectral index, $\alpha_s \equiv d n_s / d {\rm ln} k$, which affects the scalar
power spectrum as follows:
\begin{equation}\label{powerspectrum}
P(k) = A \, {\rm exp}\Big[ (n_s -1){\rm ln}(k/k_0) + \frac{1}{2} \alpha_s \, {\rm ln}^2 (k/k_0)\Big]~,
\end{equation}
where $k_0$ is a pivot point, typically taken to have the value $k_0=0.002$: 
see~\cite{wmap,Planck}. The parameter $\alpha_s$ is
given in terms of the effective inflationary potential and the slow-roll parameters by~\cite{rsi}
\begin{eqnarray}
\alpha_s & = &
- \frac{1}{32\pi^2}\left({M_{Pl}}^3\frac{V^{'''}}{V}\right) \left(M_{Pl} \frac{V'}{V}\right)
+ \frac{1}{8\pi^2} \left( M_{Pl}^2 \frac{V^{''}}{V}\right) \left(M_{Pl} \frac{V'}{V}\right)^2
- \frac{3}{32\pi^2}\left(M_{Pl} \frac{V'}{V}\right)^4 \nonumber \\
& = & \frac{1}{8\, \pi^2} \Big[ - \frac{\xi}{4} + 2 \, \eta\,  \epsilon - 3\, \epsilon^2 \Big] .
\label{alphas}
\end{eqnarray}
This is in principle an important ambiguity in fits to the CMB data:
for example, the general inflationary fit to the Planck data yields $\alpha_s = - 0.0134 \pm 0.0090$~\cite{Planck},
which is compatible with zero at the 1.5-$\sigma$ level.
However, $\alpha_s$ is expected to be very small in generic slow-roll models. Here we verify
our models indeed predict that $\alpha_s$ is small, so that the predictions of  (\ref{model}, \ref{WZW}))
can be confronted with the data assuming that $\alpha_s \simeq 0$.

\section{Application to the Single-Field Model}

\begin{figure}[ht]
\centering \includegraphics[width=10cm]{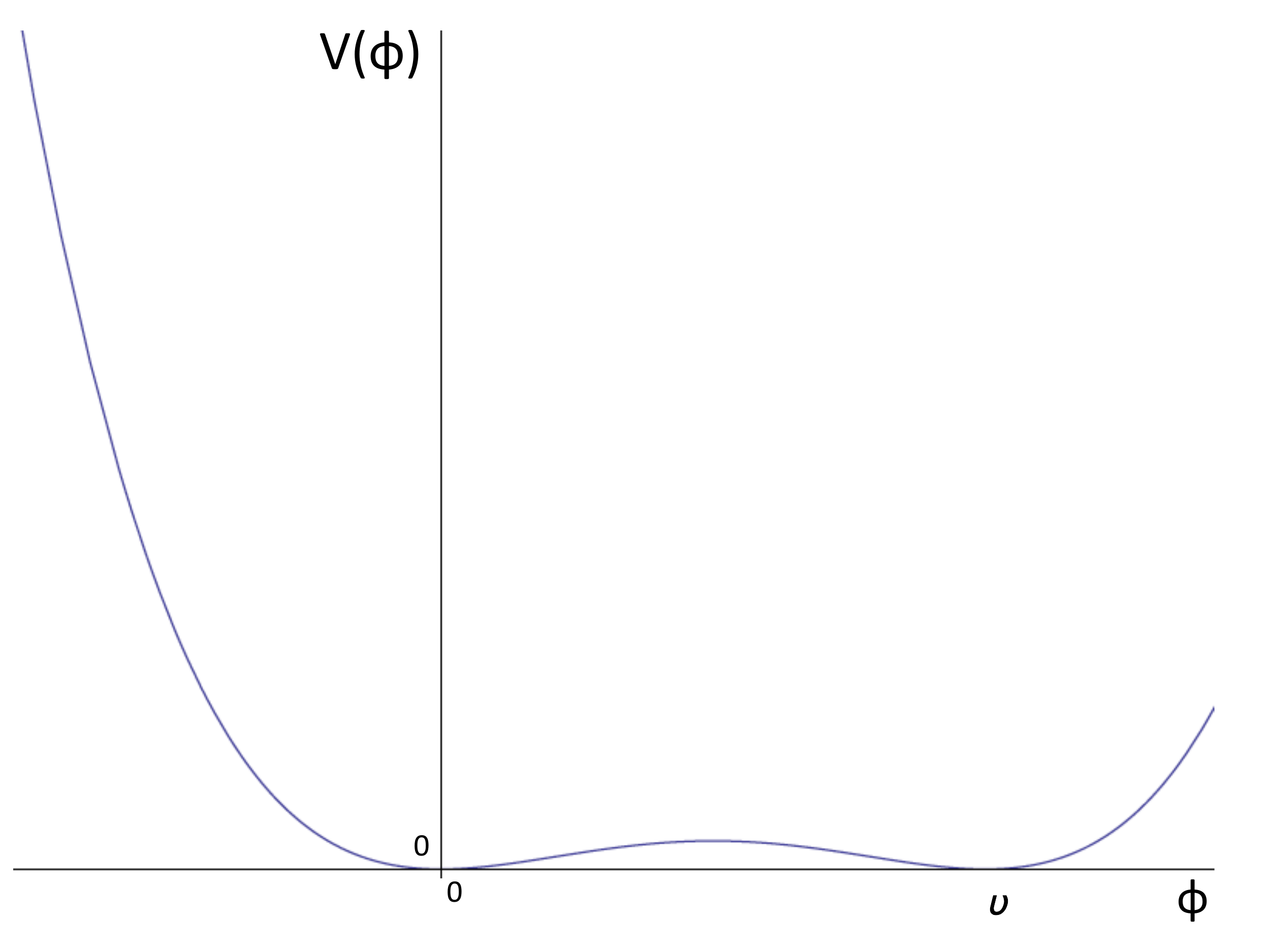} \caption{\it
The shape of the effective potential (\ref{model}) of the minimal single-field
model.}
\label{fig:simple}
\end{figure}

The potential of the minimal single-field model (\ref{model}) is displayed in Fig.~\ref{fig:simple}.
The only one of the equations in the previous Section that is inhomogeneous in $A$,
or equivalently $\lambda$, is that for the overall magnitude of the
density perturbations (\ref{WMAPcon}), so this can be used to
fix the value of $A$ ($\lambda$) following the rest of the analysis. The magnitude of $v$ is fixed as a
function of $x_i$ by the number of e-folds $N$ (\ref{Nformula}), and is
$\gg M_{Pl}$ for any value of $x_i$, as seen in the Table for $N = 50$ and some representative
values of $x_i$. Hence the slow-roll conditions
$\epsilon, \eta, \xi \ll 1$ are always satisfied and $\alpha_s$ is always negligible,
as seen in the penultimate row of the Table.

\begin{table}[htb]
\begin{center}
\vspace{0.5cm}
\begin{tabular}{|c||c|c|c|c|}
\hline\hline
Value of $x_i$ & 0.1 & 0.2 & 0.3 & 0.4 \\
\hline
Derived quantity &  &  &  &   \\
\hline\hline
$\frac{v^2}{M_{Pl}^2}$ & $18000$ &  4200 & 1600 & 710 \\ 
$\epsilon$ & 0.0085 &  0.0067 & 0.0045 & 0.0020 \\
$\eta$ &0.0062 &  0.00074 & -0.0073 & -0.022 \\
$\xi$ & -0.000053   & -0.000077  & -0.000079 & -0.000050 \\
\hline
$r$ & 0.14 & 0.11 & 0.072 & 0.031 \\
$n_s$ & 0.961 &  0.961 & 0.958 & 0.945 \\
$\alpha_s$ & $-1.4 \times 10^{-6} $ &  $-1.3 \times 10^{-6} $ &  $-1.4 \times 10^{-6} $ &  $-1.1 \times 10^{-6} $ \\
$\lambda$ & $4.3 \times 10^{-8}$ & $1.0 \times 10^{-7}$  & $2.1 \times 10^{-7} $ & $4.1 \times 10^{-7}$  \\
\hline
\end{tabular}
\end{center}
\caption{\it Numerical predictions in the simplified model (\ref{model}) for representative
values of $x_i$ and calculated for $N = 50$, showing that
$v \gg M_{Pl}$, that $\epsilon, \eta, \xi \ll 1$, that $\alpha_s$ is negligible,
and that $r$ and $n_s$ are both compatible with the WMAP data for
$0.2 \la x_i \la 0.3$.}
\label{tab:JEnumbers}
\end{table}

In the limit $x_i \to 0^\pm$ we
recover the standard predictions of $\phi^2$ models, including a value for
$r \sim 0.15$ that was only marginally compatible with the WMAP data~\cite{wmap} 
and is strongly disfavoured by the Planck data~\cite{Planck}. As seen in Fig.~\ref{fig:simple}, the potential rises more
rapidly than $\phi^2$ for $x < 0$, so negative values of $x_i$ would 
yield larger values of $r$, increasing towards
the standard predictions for $\phi^4$ models for large negative $x_i$, 
which are now very strongly excluded~\cite{Planck}.

The situation is completely different for $x_i \to 1/2^-$, as seen in Fig.~\ref{fig:simple} and
the Table. Since the potential rises much less rapidly than the $\phi^2$ case in this region,
we find that $\epsilon$ decreases monotonically as $x_i \to 1/2^-$,
and consequently that $r$ may be much smaller than in the $\phi^2$ model,
and {\it a fortiori} also the $\phi^4$ model. We also see that $\eta$ decreases
as $x_i$ increases, passing
through zero and becoming negative for $x_i \ga 0.21$. This reflects the fact that the
curvature of the potential $\propto V''$ changes from being positive in the neighbourhood
of the minimum at $x = 0$ to being negative in the neighbourhood of the local maximum at
$x = 0.5$. As a consequence, $n_s$ decreases as $x_i \to 0.5^-$, becoming smaller than
the preferred experimental range when $x_i \ga 0.4$, if $N = 50$. However, we emphasize that the value of $n_s$
is sensitive to the number of e-folds assumed, that the numbers in the Table are calculated for $N = 50$,
and that larger values of $N$ would yield values of $n_s$ closer to unity. The Table
shows that the simplified model (\ref{model}) gives acceptable inflation for $x_i \ga 0.2$.

\begin{figure}[ht]
\centering \includegraphics[width=10cm]{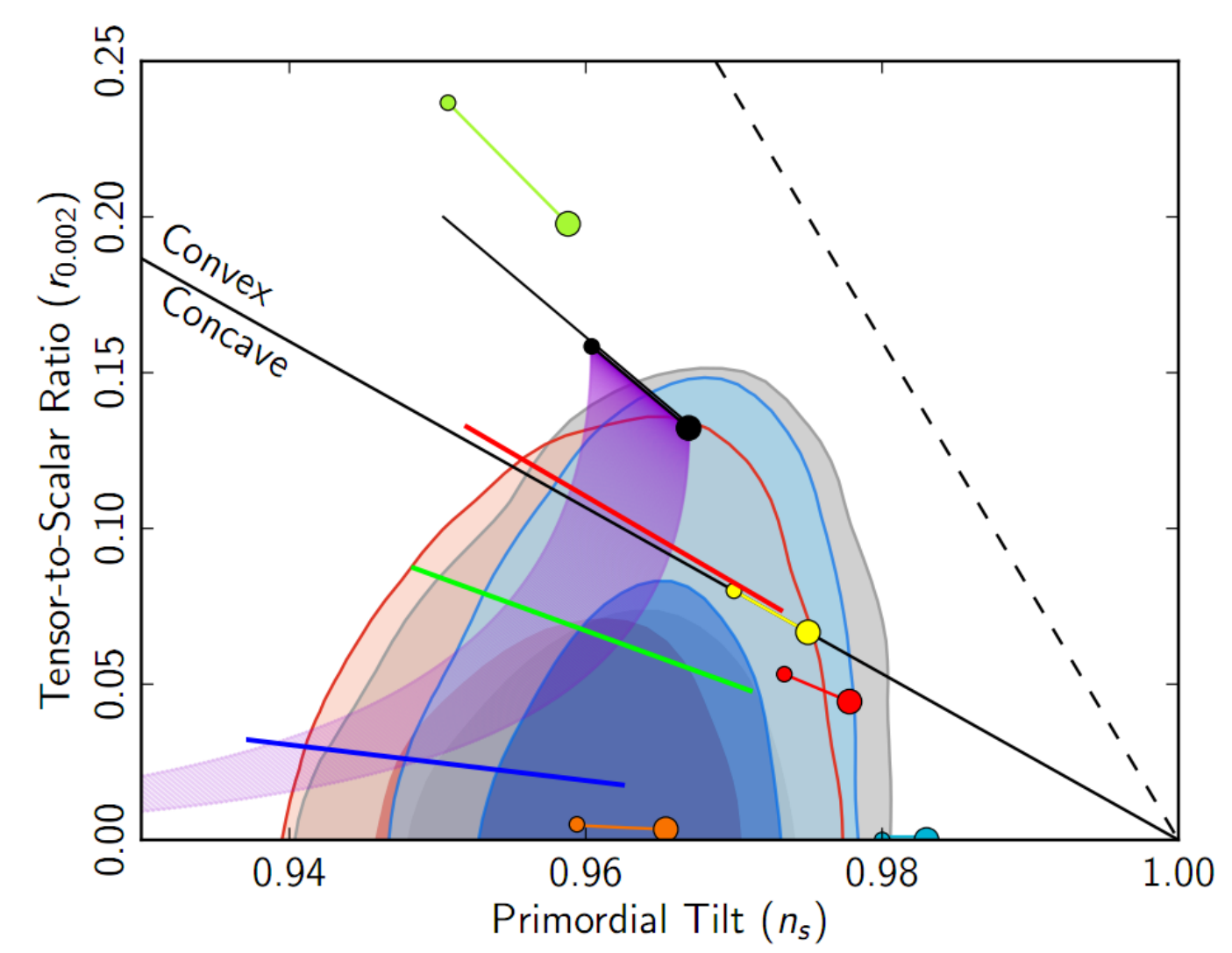} \caption{\it
Predictions in the $(n_s, r)$ plane
of our model for inflation, based on an inflation potential of the form (\ref{model}, \ref{xparam})
for various values of $x_i$: 0.2 (red), 0.3 (green) and 0.4 (blue) in the range $40 < N < 70$,
compared with the Planck constraints~\cite{Planck}. Also shown are the predictions of various other
models for inflation in the range $50 < N < 60$, also taken from~\cite{Planck}.}
\label{fig:planckcomp}
\end{figure}

The predictions of the single-field model (\ref{model}, \ref{xparam}) are displayed more completely in
Fig.~\ref{fig:planckcomp}, where they are also compared with the Planck constraints~\cite{Planck}.
We see that the model predictions enter well within the Planck 95\% CL region in the $(n_s, r)$ plane
for most of the range $40 < N < 70$ for $x_i \ge 0.2$. In contrast, the predictions of the $\phi^2$
model barely graze the 95\% CL region for $60 \la N \la 70$. Even worse are other simple inflationary models
with monomial $\phi^n: n > 2$ potentials: only the potentials $\propto \phi$~\cite{MSW,Hiranya}
and $\phi^{2/3}$~\cite{SW} enter within the Planck 95\% CL range~\cite{Planck}.

Before leaving the simple model (\ref{model}), we comment on the possibility
of topological inflation in this scenario. Since this model has two distinct vacua
with $\phi = 0, v$ that have zero energy, one could imagine that the pre-inflationary
dynamics would populate the Universe roughly equally with regions of these vacua,
separated by domain walls. As pointed out in~\cite{Linde,Vilenkin}, under certain
conditions the domain walls between these regions could inflate. The numerical
conditions for successful topological domain wall inflation were explored in~\cite{Sakai},
with the conclusion that the constraint $v \ga 0.16 M_{Pl}$ would suffice, independent of
$\lambda$~\footnote{The papers~\cite{Linde,Vilenkin,Sakai} considered models with
$V(\varphi) = \lambda (\varphi^2 - {\hat v}^2)^2$, which are seen to be equivalent to (\ref{model})
when one identifies ${\hat v} = v/2$ and $\varphi = \phi - v/2$.}. It is clear from the estimates of
$v$ in the Table that the condition found in~\cite{Sakai} is comfortably satisfied in the model
(\ref{model}).

\section{Extension to the Wess-Zumino Model}

\begin{figure}[ht]
\centering \includegraphics[width=10cm]{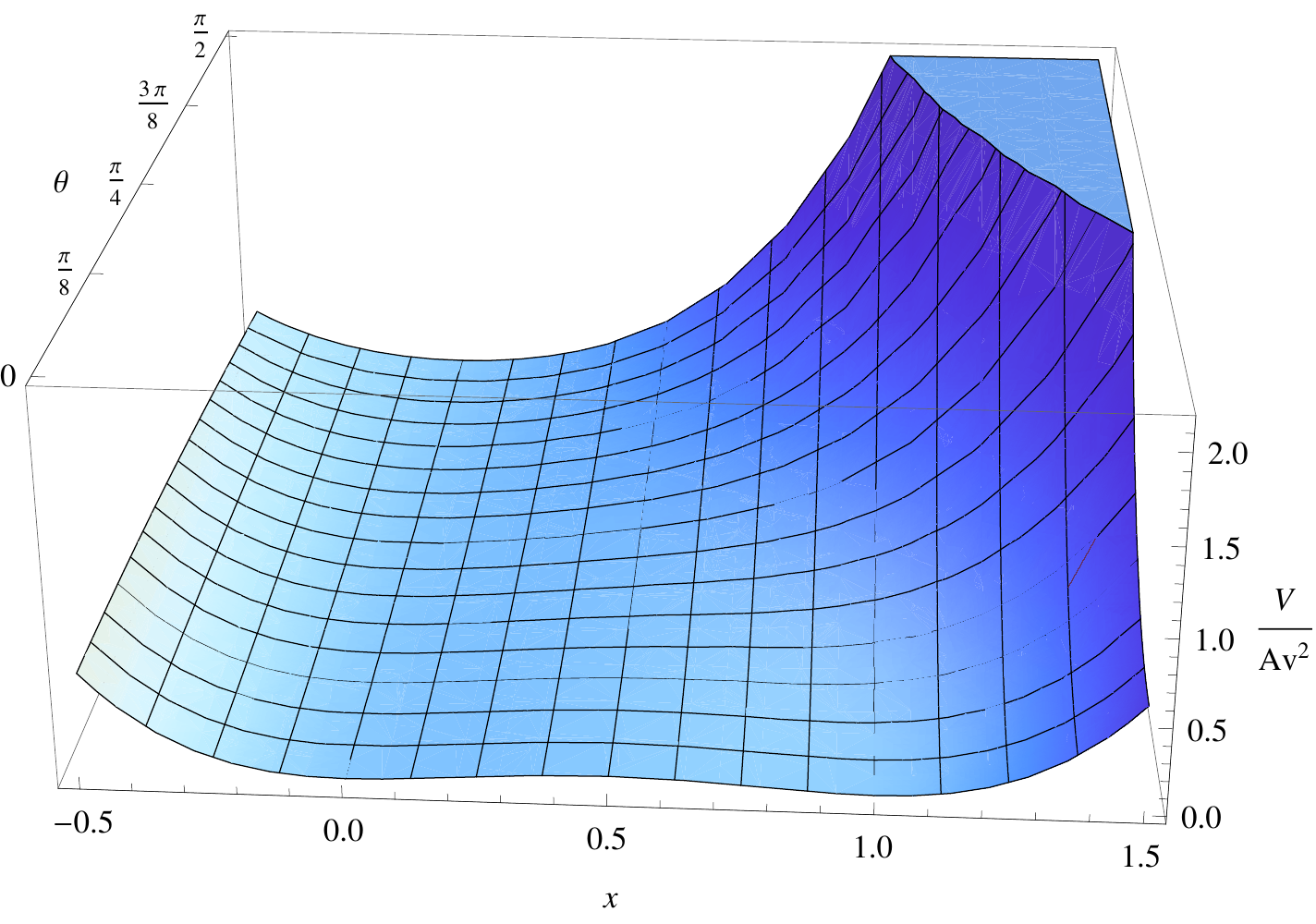} \caption{\it
The shape of the effective potential (\ref{xparam}) of the single-superfield
Wess-Zumino model (\ref{WZW}) as a function of $x$ and $\cos \theta$
for $0 \le x \le 0.5$ and $\cos \theta \ge 0$.}
\label{fig:WZW}
\end{figure}

We now proceed to the one-superfield Wess-Zumino model characterized by the
effective potential (\ref{WZW}) in which the additional degree of freedom
parameterized by $\theta$ appears as in (\ref{xparam}). It is clear that there is an equivalence
between the configurations $(\cos \theta, x) \leftrightarrow - (\cos \theta, x)$,
so we restrict our attention here to the portion of parameter space with $\cos \theta \ge 0$.
Fig.~\ref{fig:WZW} displays the effective potential (\ref{xparam}) in this region.
When $\cos \theta$ is small, the only minimum of the potential (\ref{xparam}) is that
with $x = 0$. A second, local minimum develops only for $\cos \theta > \sqrt{8/9}$, but
this has positive energy, falling to zero only when $\cos \theta \to 0$. 

Along the boundary where $\cos \theta = 1$, the form of the effective potential
(\ref{xparam}) is identical to that in the single-field model (\ref{model}), and the
discussion of inflation given in the previous Section goes through unchanged.
On the other hand, the potential (\ref{xparam}) vanishes along the boundary $x = 0$.
At any fixed positive value of $x \ne 0$, the potential increases monotonically as
$\cos \theta$ decreases from $1 \to 0^+$. In particular, when $\cos \theta = 0$ ($\theta = \pi/2$), 
the potential is a combination of quadratic and quartic terms with coefficients
of the same sign, a scenario that is excluded by the CMB data~\cite{wmap,Planck}. 
A complete discussion of the inflationary
possibilities for initial conditions at arbitrary points in the $(x \cos \theta)$ plane
lies beyond the scope of this work, but it is clear that, although successful inflation cannot be
obtained when $\cos \theta = 0$, it would be possible in a neighbourhood of $\cos \theta = 1$.

\section{Combination with the Seesaw Model of Neutrino Masses}

We now discuss how such a Wess-Zumino inflationary model could be
combined with the minimal supersymmetric seesaw model. In this case,
one would identify the superfield $\Phi$ with the singlet (right-handed)
sneutrino superfield. In this case, the quadratic term in (\ref{WZW}) would
generate $\Delta L = 2$ processes (where $L$ is lepton number), corresponding
to a Majorana neutrino mass. These processes would conserve $R$ parity.
On the other hand, the trilinear term in (\ref{WZW}) would generate $\Delta L = 3$
processes, which would violate $R$ parity and cause the lightest supersymmetric
particle (LSP) to be unstable, in general. However, the rate of $R$ violation
would be very small, so the LSP could still provide the astrophysical cold dark matter.

Consider, for example, the case in which the LSP is the gravitino ${\tilde G}$. This would have a
tree-level coupling to a singlet antisneutrino-neutrino pair. The singlet neutrino
would mix with the conventional left-handed neutrino via a Yukawa vertex with
a Standard Model Higgs scalar vacuum expectation value divided by the large singlet neutrino mass. On the other hand,
the singlet antisneutrino would couple via the the trilinear coupling in (\ref{WZW})
to a pair of singlet neutrinos, which would also mix with left-handed neutrinos.
This and similar diagrams would give rise to ${\tilde G} \to 3 \nu$ decay, but at
a very low rate, suppressed by several factors of the heavy singlet-neutrino
mass scale.

\section{Conclusions}

The very precise Planck data~\cite{Planck} are generally consistent with the
idea of cosmological inflation ({\it modulo} a few well-publicized anomalies),
but pose considerable challenges for simple inflationary models. Indeed, no
single-field model with a monomial potential $\propto \phi^n: n \ge 2$ is
comfortably consistent with the data. However, we have shown in this paper
that a simple single-field model of the form (\ref{model}) is highly consistent
with the data. Moreover, we have shown that this potential arises very naturally
within the simplest single-superfield Wess-Zumino model (\ref{WZW}). Finally, we
have also shown that this model may be combined with a minimal supersymmetric
seesaw model of neutrino masses.

The most important pressure on this model comes from the Planck upper limit
on the tensor-to-scalar ratio $r$, and we look forward to future improved
constraints on this quantity from CMB polarization data from Planck and other
experiments. If the upper limit on $r$ were to be reduced significantly, this
would favour variants of the model with larger values of $x_i \to 0.5^-$, in which case
the model might be consistent with the observational constraint on $n_s$
for only a more restricted range of $N$.

In the mean time, it would be interesting to explore in more detail the possible
predictions of the Wess-Zumino model (\ref{WZW}) for $\cos \theta > 0$, the
possibility of topological inflation, and possible observational signatures of
the small violation of $R$ parity that this model would predict if combined
with a supersymmetric seesaw model of neutrino masses.

\section*{Acknowledgements}

The work of JE and NEM was supported partly by the London
Centre for Terauniverse Studies (LCTS), using funding from the European
Research Council via the Advanced Investigator Grant 267352.
They thank CERN for kind hospitality.

\end{document}